\documentclass{article}
\usepackage[left=3cm,right=3cm,top=3cm,bottom=3cm]{geometry} 
\usepackage{amsmath} 
\usepackage{authblk}
\providecommand{\keywords}[1]{\textbf{\textit{Keywords:}} #1}

\usepackage{graphicx}
\usepackage{mathtools}
\mathtoolsset{showonlyrefs=true}
\usepackage{natbib}
\usepackage{amssymb}
\usepackage{amsmath}
\usepackage{float}
\usepackage{subfig}
\usepackage{xcolor}
\usepackage{url}
\begin{document}

\title{The wrapped skew Gaussian process for analyzing spatio-temporal data
}


\author[1]{Gianluca Mastrantonio}
\author[2]{Giovanna Jona Lasinio}
\author[3]{Alan E. Gelfand}

\affil[1]{ Roma Tre University, Via Silvio D'Amico 77,  Rome, 00145, Italy}
\affil[2]{Sapienza University of Rome, P.le Aldo Moro 5,  Rome, 00185, Italy}
\affil[3]{Duke University,  223-A Old Chemistry Building Box 90251,  Durham, NC 27708-0251, USA}

\date{}

\maketitle

\begin{abstract}
	We consider modeling of angular or directional data viewed as a linear variable wrapped onto a unit circle.  In particular, we
focus on the spatio-temporal context, motivated by a collection of wave directions obtained as computer model output developed
	dynamically over a collection of spatial locations. We propose a novel wrapped skew Gaussian process which enriches the class of
	wrapped Gaussian process. The wrapped skew Gaussian process  enables more flexible marginal distributions than the symmetric ones
	arising under the wrapped Gaussian process and it allows straightforward interpretation of parameters.  We clarify that replication through time enables criticism of
	the wrapped process in favor of the wrapped skew process.\\	
	We formulate a hierarchical model incorporating this process and show how to introduce appropriate latent variables in order to enable
	efficient fitting to dynamic spatial directional data. We also show how to implement kriging and forecasting under this model.
	We provide a simulation example as a proof of concept as well as a real data example. Both examples reveal consequential improvement in predictive
	performance for the wrapped skew Gaussian specification compared with the earlier wrapped Gaussian version.


\end{abstract}

\section{Introduction}
There is increasing interest in analyzing directional data which are collected over space and time.  Examples arise, for
instance, in oceanography (wave directions), meteorology (wind directions), biology (study of animal movement).  They also arise
from periodic data, e.g., event times might be wrapped according to a daily period to give a circular view (eliminating
\emph{end} effects). We wrap time around a circle by a modulus transformation and, without loss of generality, can rescale to degrees or angles on a unit circle.  Time wrapping with spatial data occurs naturally in applications such as locations and times of crime events, locations and times of automobile accidents, and residence address with time of admission for hospitalizations.

\cite{Jona2013} introduced a Bayesian hierarchical model to handle angular data, enabling full inference regarding all model
parameters and prediction under the model.  Their context was multivariate directional observations arising as angular data
measurements taken at spatial locations, anticipating structured dependence between these measurements. They proposed the
wrapped spatial Gaussian process, induced from a linear spatial Gaussian process. They explored dependence structure and showed
how to implement kriging of mean directions and concentrations in this setting.

The current state of the art for modeling circular space-time data includes the wrapped Gaussian process and the projected Gaussian process.  The second, although more flexible, is based upon a four parameter model such that complex interactions among the parameters make interpretation difficult.
In this paper our contribution is to overcome a key limitation of the wrapped  Gaussian process, that the marginal distributions at all
locations are symmetric. Here we introduce the wrapped skew
Gaussian process.  This new circular process allows for asymmetric marginal distributions while retaining straightforward parametric interpretation.
Our wrapping approach is developed from
the skew normal distribution proposed by \cite{Azzalini:1985} and the process extension constructed by
\cite{Zhang2010}\\

By now, there is a fairly rich literature on skew
multivariate normal models \citep{azzalini2005,sahu2003,ma2004,wang2004} but all are \emph{inline}, i.e., on a linear scale.
The first attempt to wrap the skew normal distribution for circular data can
be found in \cite{Pewsey2000} where its basic properties are derived. Follow-on work appears in \cite{Pewsey2006,hernandez2012}.

To our knowledge, we propose the first extension to multivariate wrapped skew distributions, in
particular, to a spatial and spatio-temporal setting. In what follows we review the univariate wrapped skew normal distribution, showing the flexibility of shapes and do the same for bivariate wrapped
skew normal distributions.  Then, we turn to a hierarchical model for dynamic spatial data and show how, using suitable latent variables, to fit it efficiently. We also show how to implement kriging under this model.

%

A critical point emerges: though we can fit both models with a single sample of spatially referenced directions, in terms of kriging performance, we can not criticize the wrapped spatial Gaussian process in favor of the wrapped skew spatial Gaussian process.  This is not surprising.
Consider the linear situation. With a single sample of data from a set of locations, it is difficult to criticize the Gaussian
process in favor of a more complex stochastic process specification, i.e., it is difficult to criticize a multivariate  normal
model with a single sample of multivariate data.  However, with replicates, we are able to demonstrate substantially improved
predictive performance for the wrapped skew Gaussian process.  We do this both with simulated data, as a proof of concept, and
with real data, making direct comparison. In our setting replicates arise through a dynamic spatial data where we envision i.i.d. spatial increment processes.

Inference for spatial data is challenging due to the restriction of support to the unit circle, $[0,2\pi)$, and to the
sensitivity of descriptive and inferential results to the starting point on the circle. There exists a substantial early literature on
circular data (see e.g.  \cite{mardia72} and  \cite{Merdia1999}, \cite{Jammalamadaka2001}  or  \cite{fisher1996}) primarily confined to descriptive statistics and limited inference for simple univariate models.

Computational procedures such as MCMC methods and the EM
algorithm, have substantially advanced inference opportunities for directional data. Some examples include linear models
\citep[][]{harrison,fisher1996,Katoa2008}, linear models in a Bayesian context \citep[][]{guttorp1998, damien-walker}, models
for circular time series \citep[][]{breckling1989, coles98, Merdia1999, ravindran, hughes, Fisher1992, Holzmann2006} or model for space-time circular-linear data \citep{lagona2015}.
Recently,
\cite{kato2010}, building upon earlier work \citep{kato2008}, proposed a discrete time Markov process for circular
data. He  uses the M\"{o}bius circle transformation, connecting it with an early Markov process model of \cite{fisher1994}.

With regard to multivariate theory for circular data, particularly in the fully Bayesian setting, the work of \cite{coles98} is
foundational for ours.  He also employs wrapped distributions, noting that, in the Gaussian case, they can be readily given a
multivariate extension.  Coles mostly works with independent replicates of multivariate circular data in low dimension with an
unknown covariance matrix and develops some theory and examples for the time series setting. He mentions possible extensions to
the spatial setting but offers no development, in particular, no thoughts on regression or kriging (Sections 3.5 and 3.6 below).
\citet{Coles1998} include spatial dependence in looking at the direction of maximum wind speed.  With little detail, they
propose conditionally independent directions modeled with a von Mises distribution, introducing spatial structure in the modal
direction and concentration parameters, a second stage specification. Our view, again following \citet{Jona2013}, is to
introduce spatial structure at the first stage of the modeling, directly on the angular variables, resulting in a spatial
process model with smooth process realizations.

Following a different strand, the projected normal and the
associated projected Gaussian process \citep{Wang2013,wang2014} have generated recent interest.  In particular,  a
general bivariate normal distribution is projected to an angle, extending work of \cite{Presnell1998}   and    \cite{nunez2005}.  The extension to a stochastic process for variables on the circle over a continuous spatial domain, the
\emph{projected Gaussian process}, is induced from a linear bivariate spatial Gaussian process. The projected Gaussian process
has marginal distributions that can be asymmetric, possibly bimodal, an advantage over the wrapped Gaussian process.
\citet{wang2014} also investigate properties of this process, including the nature of joint distributions
for pairs of directions at different locations.  Working within a hierarchical Bayesian framework, they show that model fitting
is straightforward using suitable latent variable augmentation in the context of Markov chain Monte Carlo (MCMC).  In very
recent work, \cite{mastrantonio2015b}  offer comparison between the wrapping and
the projection modeling approaches.

We remark that we have explored the possibility of introducing skewness into the projected Gaussian process.  The overall
process model is induced by a bivariate skewed Gaussian process.  This is a more challenging process to work with; the resulting directional process model is extremely messy and
has proved very difficult to fit. It likely exceeds what the data is capable of supporting. We do not discuss it further.

Our motivating example is drawn from marine data.  Wave heights and outgoing wave directions, the latter being measured in
degrees relative to a fixed orientation, are the main outputs of marine forecasts. Numerical models for weather and marine
forecasts need  statistical post-processing. Wave directions, being angular variables, cannot be treated through standard
post-processing techniques \citep[see][and references therein]{engel2007,bao2009}.  In \citet{bao2009}  bias correction and  ensemble
calibration forecasts of surface wind direction are proposed. The authors use circular-circular regression as in \citet{kato2008}
for bias correction and  Bayesian model averaging with the von Mises distribution for ensemble calibration.  However, their
approach does not explicitly account for spatial structure.

Lastly, it is worth commenting that, in our setting, wave direction data is viewed differently from wind
direction data.  The former is only available as an angle while the latter is customarily associated with wind speed, emerging
as the resultant of North-South and East-West wind speed components.


The format of the paper is as follows.  In Section 2 we review, develop and illustrate the univariate wrapped skew normal
distribution.  Section 3 extends to the wrapped skew Gaussian process, including distribution theory, model fitting, and
kriging.  Section 4 provides the dynamic version which we then pursue through simulation in Section 5 and a wave direction data
analysis in Section 6.  Section 7 offers a brief summary and some future research possibilities.

\section{The  wrapped skew normal}

\subsection{The univariate case} \label{sec:unv}

We begin with the univariate wrapped skew normal distribution. Let $X$ and $W$ be two independent
standard normal variables, let $\sigma^2 \in \mathbb{R}^+$ and $\lambda \in \mathbb{R}$. Then, the random variable


\begin{equation} \label{eq:zeta}
	Z  = \mu+  \frac{\sigma \lambda}{\sqrt{1+\lambda^2}  }   |X|+  \frac{\sigma}{\sqrt{1+\lambda^2}  } W-  \frac{\sigma \lambda \sqrt{ 2  }}{ \sqrt{\pi(1+\lambda^2)}  }
\end{equation}
is said to be distributed as a skew normal variable \citep{Azzalini:1985} with parameters $\mu$, $\sigma^2$ and $\lambda$; i.e., $Z| \boldsymbol{\Psi} \sim SN(\mu, \sigma^2, \lambda)$, where  $ \boldsymbol{\Psi}$ denotes the vector of parameters. Let $\phi(\cdot)$ and $\Phi(\cdot)$ be the probability density
function (pdf) and the cumulative density function (cdf), respectively, of a standard normal.  Then,
the pdf of $Z|\boldsymbol{\Psi}$ is
\begin{equation} \label{eq_densz}
	\frac{2}{\sigma}\phi\left( \frac{z-\mu+  \frac{\sigma \lambda \sqrt{ 2  }}{ \sqrt{\pi(1+\lambda^2)}  }   }{\sigma} \right)\Phi \left( \lambda \left( \frac{z-\mu+ \frac{\sigma \lambda \sqrt{ 2  }}{ \sqrt{\pi(1+\lambda^2)}  } }{\sigma} \right)  \right)
\end{equation}
and from \eqref{eq:zeta} we can easily derive  the mean and the variance of $Z$, respectively. They are  $\mu$ (the definition in (\ref{eq:zeta}) was made in order to center $Z$ at $\mu$) and $${\sigma^2 \lambda^2}/{( {1+\lambda^2})  }  \left(1-{2}/{\pi}\right)+ {\sigma^2 }/{ ({1+\lambda^2})  }.$$


With the transformation
\begin{equation} \label{eq:wrap}
	\Theta = Z \mbox{ mod } 2 \pi, \hspace{.2cm} \mbox{implying } \Theta \in [0,2\pi), 
\end{equation}
we obtain a random variable with support on the unit circle. We can express the inline variable as $Z = \Theta+2 \pi K$, where $K $, the \emph{winding number}, assumes values in $ \mathbb{Z}= \{0,\pm1, \pm2,...\}$.  The transformation \eqref{eq:wrap} defines what is called  a \emph{wrapped skew normal} (WSN) distribution, as introduced in  \cite{Pewsey2000}.  It wraps the skew normal distribution, defined on the real line, onto the unit circle. Details on the wrapping approach
can be found  in \cite{Jammalamadaka2001} or \cite{Merdia1999}. 

The pdf of $\Theta|\boldsymbol{\Psi}$ is
\begin{align}
	\sum_{k \in \mathbb{Z}} \frac{2}{\sigma} &\phi\left( \frac{\theta+2 \pi k-\mu+  \frac{\sigma \lambda \sqrt{ 2  }}{ \sqrt{\pi(1+\lambda^2)}  }       }{\sigma} \right)\\
	&\times \Phi \left( \lambda \left( \frac{\theta+2 \pi k-\mu+\frac{\sigma \lambda \sqrt{ 2  }}{ \sqrt{\pi(1+\lambda^2)}  } }{\sigma} \right)  \right).\label{eq:x}
\end{align}
The infinite sum in \eqref{eq:x} is impossible to evaluate but, to display the density, as with the wrapped normal case, we can
obtain an accurate approximation by appropriately truncating the sum.  Figure 1 illustrates the effect of introduction of skewness into the wrapped normal density.  To obtain a sample from a wrapped skew normal we first
obtain a sample from the skew normal and then transform it to a circular variable via
\eqref{eq:wrap}.  Also, note that, if we let $K$  be a random variable, the density inside the sum in \eqref{eq:x} is the joint density of
$(\Theta,K|\boldsymbol{\Psi})$ whence, we marginalize over $K$ to obtain the density of the circular variable.

\begin{figure}[t!]
	\centering
	\caption{Densities of the  wrapped skew normal (solid line) with $\mu= \pi$, $\sigma^2=1$ and different values of  $\lambda$  along with the associated densities of the wrapped normal (dashed line) having the same circular mean and variance } \label{fig:skew1}
	\subfloat[ $\lambda= 3$]{\includegraphics[trim=50 120 20 80,clip,scale=0.3]{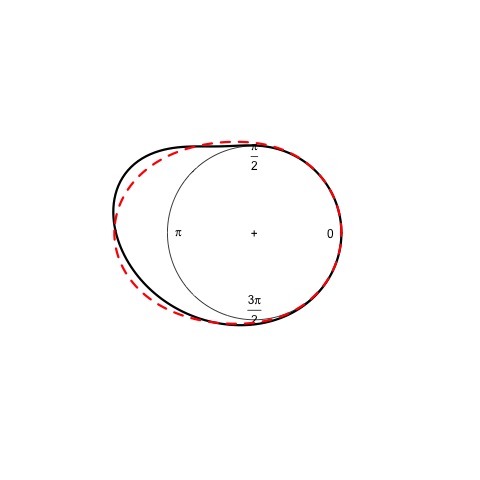}}
	\subfloat[  $\lambda= 10$]{\includegraphics[trim=50 120 20 80,clip,scale=0.3]{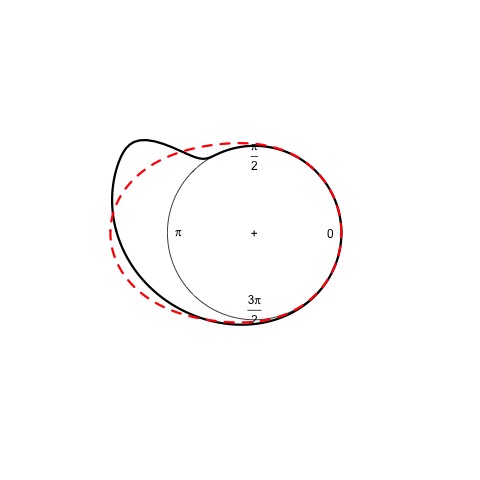}}\\
	\subfloat[  $\lambda= -10$]{\includegraphics[trim=50 120 20 80,clip,scale=0.3]{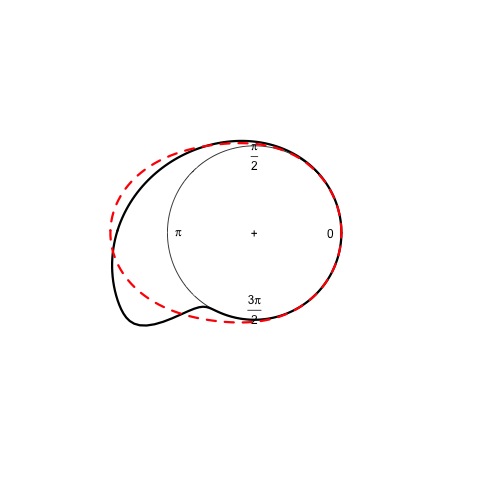}}
\end{figure}

\cite{Pewsey2000} gives the fundamental properties of the  WSN along with closed forms for  the cosine and sine moments. Let $\mu^* = \mu-\frac{\sigma \lambda \sqrt{ 2  }}{ \sqrt{\pi(1+\lambda^2)}} $ and $\mathcal{J}(a) = \int_{0}^a \sqrt{\frac{2}{\pi}} \exp\left( \frac{u^2}{2} \right)d u$ the cosine and sine moments
become
\begin{align}
	\alpha_p &= E(\cos p \Theta|\boldsymbol{\Psi}) = \exp\left( -\frac{{p^2}\sigma^2 }{2}    \right)\\
	&\times
	\left(    \cos (p \mu^*) - \mathcal{J}\left( \frac{\lambda \sigma p}{\sqrt{1+\lambda^2}} \right)\sin(p \mu^*)   \right) \label{eq:cos}
\end{align}
and
\begin{align}
	\beta_p &= E(\sin p \Theta|\boldsymbol{\Psi}) =
	\exp\left( -\frac{{p^2}\sigma^2 }{2}    \right) \\
&\times	\left(    \sin (p \mu^*) +\mathcal{J}\left( \frac{\lambda \sigma p}{\sqrt{1+\lambda^2}} \right)\cos(p \mu^*)   \right). \label{eq:sin}
\end{align}
The trigonometric moments \eqref{eq:cos} and \eqref{eq:sin} are useful to compute  the circular mean of $\Theta$, $\tilde{\mu} =
\mbox{atan}^*\frac{\alpha_1}{\beta_1}\footnote{For the definition of $\mbox{atan}^*$ see \cite{Jammalamadaka2001}, p. 13}$,
and the circular concentration, $\tilde{c} = \sqrt{\alpha_1^2+\beta_1^2}$.  However, unfortunately we need to compute
$\mathcal{J}(\cdot)$, which is not available in closed form.  \cite{Pewsey2000} suggests to use deterministic
numerical integration methods but we note that  $\alpha_p$ and $\beta_p$ can  be computed using Monte Carlo approximation.

Indeed, from \eqref{eq:zeta} we can see that
\begin{equation}\label{eq:cond}
	Z | X,\boldsymbol{\Psi} \sim N\left(\mu+ \frac{\sigma \lambda}{\sqrt{1+\lambda^2}} |X|- \frac{\sigma \lambda \sqrt{ 2  }}{ \sqrt{\pi(1+\lambda^2)}}  ,
	\frac{\sigma^2}{1+\lambda^2}  \right)
\end{equation}
and as a consequence
\begin{equation}\label{eq:cond2}
	\Theta | X,\boldsymbol{\Psi} \sim WN\left(\mu+ \frac{\sigma \lambda}{\sqrt{1+\lambda^2}} |X|- \frac{\sigma \lambda \sqrt{ 2  }}{ \sqrt{\pi(1+\lambda^2)}}  ,
	\frac{\sigma^2}{1+\lambda^2}  \right),
\end{equation}
where $WN(\cdot)$ indicates the wrapped normal distribution. Let $\{X^b\}_{b=1}^B$ be a set of $B$
samples from the distribution of $X$. Then, we can write the cosine moments as $\alpha_p = E(\cos p\Theta|\boldsymbol{\Psi}) =
E_{X|\boldsymbol{\Psi}}E_{\Theta|X,\boldsymbol{\Psi}}(\cos p\Theta |X,\boldsymbol{\Psi})$, since $E_{\Theta|X,\boldsymbol{\Psi}}(\cos p\Theta |X,\boldsymbol{\Psi})  $ is the cosine moment of  $\Theta |X,\boldsymbol{\Psi}$.
Following \cite{Jona2013}, a
Monte Carlo approximation of $\alpha_p$ is
\begin{align}
	\hat{\alpha}_p &\approx \frac{\exp\left(-p^2\frac{\sigma^2}{2(1+\lambda^2)}        \right)   }{B}\\
	&\times\sum_{b=1}^B  \cos\left( p\left( \mu+ \frac{\sigma \lambda}{\sqrt{1+\lambda^2}} |X|- \frac{\sigma \lambda \sqrt{ 2  }}{ \sqrt{\pi(1+\lambda^2)}}   \right)   \right).
\end{align}
Similarly, we  find
\begin{align}
	\hat{\beta}_p &\approx \frac{\exp\left(-p^2\frac{\sigma^2}{2(1+\lambda^2)}        \right)   }{B} \\
	&\times\sum_{b=1}^B  \sin\left( p\left( \mu+ \frac{\sigma \lambda}{\sqrt{1+\lambda^2}} |X|- \frac{\sigma \lambda \sqrt{ 2  }}{ \sqrt{\pi(1+\lambda^2)}}   \right)   \right)
\end{align}
and then $\hat{\tilde{\mu}} = \mbox{atan}^*\frac{\hat{\alpha}_1}{\hat{\beta}_1} $ and $\hat{\tilde{c}} = \sqrt{\hat{\alpha}_1^2+\hat{\beta}_1^2}$.

\subsection{The bivariate case} \label{sec:multi}
Let $Z_1$ and $Z_2$ be two random variables  skew normal distributed with, respectively, parameters $(\mu_1, \sigma_1^2,\lambda_1)$ and $(\mu_2, \sigma_2^2,\lambda_2)$:
\begin{align}
	Z_1 & = \mu_1+  \frac{\sigma_1 \lambda_1}{\sqrt{1+\lambda_1^2}  }   |X_1|+  \frac{\sigma_1}{\sqrt{1+\lambda_1^2}  } W_1-  \frac{\sigma_1 \lambda_1 \sqrt{ 2  }}{ \sqrt{\pi(1+\lambda_1^2)}  }  ,\\
	Z_2 & =\mu_2+  \frac{\sigma_2 \lambda_2}{\sqrt{1+\lambda_2^2}  }   |X_2|+  \frac{\sigma_2}{\sqrt{1+\lambda_2^2}  } W_2-  \frac{\sigma_2 \lambda_2 \sqrt{ 2  }}{ \sqrt{\pi(1+\lambda_2^2)}  }  .
\end{align}
We introduce dependence between $Z_1$ and $Z_2$ by letting $\mbox{Cor}(X_1,X_2|\boldsymbol{\Psi}) = \rho_x$ and $\mbox{Cor}(W_1,W_2|\boldsymbol{\Psi}) = \rho_w$.
Then, we say that $(Z_1,Z_2|\boldsymbol{\Psi})$ is distributed as a bivariate skew normal with the additional parameters, $\rho_{x}$ and
$\rho_{w}$. This specification of the bivariate skew normal, due to \cite{Zhang2010b}, differs from the one that can be derived
using the multivariate normal of \cite{azzalini1996} and it is more suitable to built a stationary
process, see Section \ref{sec:process}.

Using the transformation \eqref{eq:wrap} we can obtain the  circular variables $\Theta_1 =  Z_1 \mbox{ mod } 2 \pi$ and
$\Theta_2 =  Z_2 \mbox{ mod } 2 \pi$ associated with $(Z_1,Z_2)$. The parameters   $\rho_x$ and $\rho_w$ govern the dependence
between $\Theta_1$ and $\Theta_2$ and  if both are 0, $\Theta_1$ and $\Theta_2$ are independent as  with the associated linear variables.

\begin{figure}[t!]
	\centering
	\caption{Bivariate densities of the  wrapped skew normal with $\mu= \pi$, $\sigma^2=1$, $\lambda=3$ in the first column and $\lambda=10$ in the second column   and several values of  $\rho_x$ and $\rho_w$. } \label{fig:bicorskew}
	\subfloat[$\rho_x= 0.2$ and $\rho_w= 0.2$]{\includegraphics[scale=0.25]{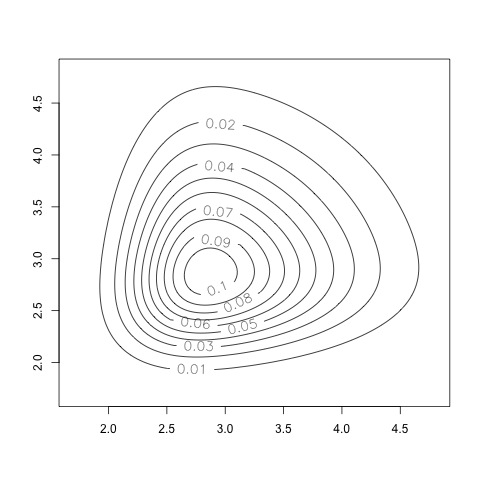}}
	\subfloat[  $\rho_x= 0.2$ and $\rho_w= 0.2$]{\includegraphics[scale=0.25]{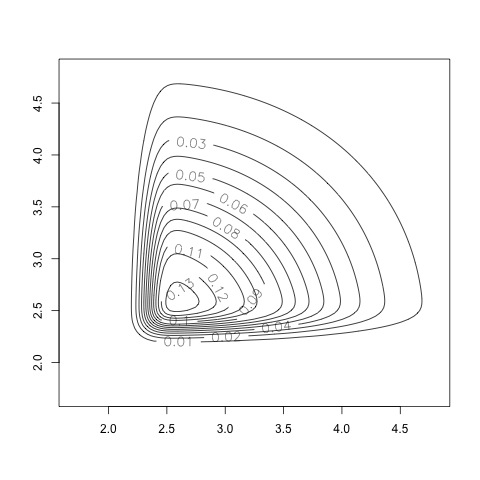}}\\
	\subfloat[   $\rho_x= 0.2$ and $\rho_w= 0.8$]{\includegraphics[scale=0.25]{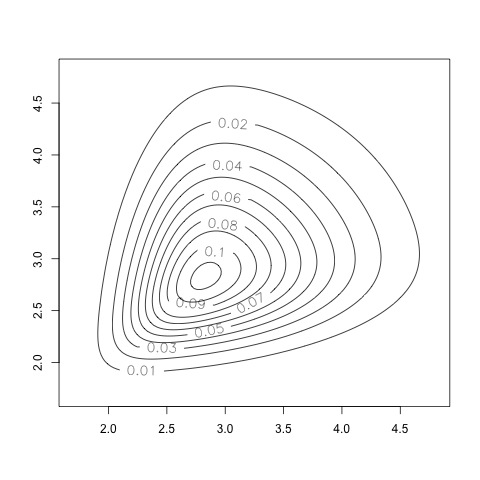}}
	\subfloat[  $\rho_x= 0.8$ and $\rho_w= 0.2$]{\includegraphics[scale=0.25]{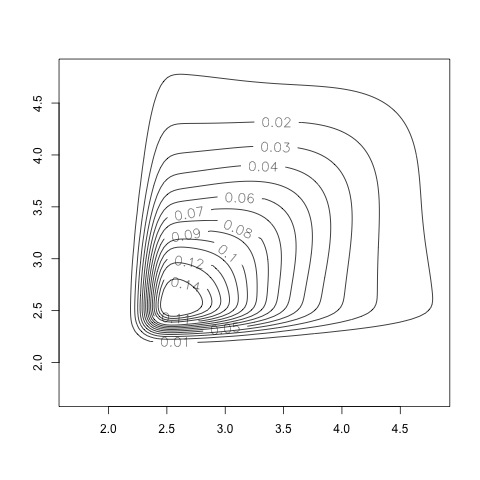}}
\end{figure}

Let $g(\cdot|\boldsymbol{\Psi})$ be the density of $(Z_1,Z_2|\boldsymbol{\Psi})^{\prime}$, let $\mathbf{K}=(K_1,K_2)^{\prime}$ be the vector of winding numbers and $\boldsymbol{\Theta} = (\Theta_1,\Theta_2)^{\prime}$, with
$\mathbf{Z}= \boldsymbol{\Theta}+2 \pi  \mathbf{K}$. As in the univariate case, we obtain the density of $\boldsymbol{\Theta}$,
a bivariate wrapped skew normal, through marginalization over $\mathbf{K}$ of the joint density of $(\boldsymbol{\Theta},
\mathbf{K}|\boldsymbol{\Psi})$:
\begin{equation} \label{eq:xmulti}
	f(\boldsymbol{\theta}|\boldsymbol{\Psi}) = \sum_{k_1 \in \mathbb{Z}} \sum_{k_2 \in \mathbb{Z}} g(\boldsymbol{\theta}+2 \pi \mathbf{k}|\boldsymbol{\Psi}).
\end{equation}
In Figure \ref{fig:bicorskew} we show  plots of the {bivariate} wrapped skew normal distributions.


%
%

\section{The  wrapped skew Gaussian process} \label{sec:process}

A natural way to construct a wrapped skew Gaussian process $\Theta(\mathbf{s}), \mathbf{s} \in \mathbb{R}^d$ is to start from a
skew Gaussian process $Z(\mathbf{s})$ on the line and define, for each $\mathbf{s}$,  $\Theta(\mathbf{s}) = Z(\mathbf{s}) \mbox{
	mod } 2 \pi$, following the approach of \cite{Jona2013}.  To capture stationarity we use the following stationary skew Gaussian
process, proposed by \cite{Zhang2010b}:
\begin{equation} \label{eq:zproc}
	Z(\mathbf{s})  = \mu+  \frac{\sigma \lambda}{\sqrt{1+\lambda^2}  }   |X(\mathbf{s})|+  \frac{\sigma}{\sqrt{1+\lambda^2}  } W(\mathbf{s})-  \frac{\sigma \lambda \sqrt{ 2  }}{ \sqrt{\pi(1+\lambda^2)}  }.
\end{equation}
Here, $X(\mathbf{s})$ and $W(\mathbf{s})$ are independent zero mean Gaussian process with isotropic parametric correlation
functions, $\boldsymbol{\rho}_x (h;\boldsymbol{\psi}_x)$ and $\boldsymbol{\rho}_w
(h;\boldsymbol{\psi}_w)$, respectively.

The process in \eqref{eq:zproc} is not the only stationary skew Gaussian process proposed in the  literature. However, \cite{Minozzo2012} point out that most of them  are in fact not stationary. For
example \cite{Kim2004} or  \cite{allard2007} built stochastic skew normal processes where the $n-$finite dimensional distributions have, as special case, the multivariate skew normal of  \cite{azzalini1999}.  But, the class of multivariate skew normal of \cite{azzalini1999} is not closed under marginalization.  Each marginal is still a skew normal but not of the same form, and  \cite{Minozzo2012} demonstrate that  the stationarity property of an $n-$dimensional finite distribution in this case is not  passed  onto the marginals.
Note that if in \eqref{eq:zproc} we let  the process $X(\mathbf{s})$ to be spatially constant, i.e.   $X(\mathbf{s}) \equiv X$,  the associated $n-$finite dimensional distributions are the \cite{azzalini1999}'s multivariate skew normal and then, from above, the process is not  stationary.
On the other hand, if the process  $W(\mathbf{s})$ is spatially constant, it is easy to demonstrate that \eqref{eq:zproc} can be written as
\begin{equation} \label{eq:zproc1}
	Z(\mathbf{s})  = \mu+  \frac{\sigma \lambda}{\sqrt{1+\lambda^2}  }   |X^*(\mathbf{s})|-  \frac{\sigma \lambda \sqrt{ 2  }}{ \sqrt{\pi(1+\lambda^2)}  },
\end{equation}
where $X^*(\mathbf{s})$ is a process with finite dimensional distributions  that are a mixture of folded normal with  mode at $\mathbf{0}$ and
covariance matrix that depends on the
covariance matrix of $X(\mathbf{s})$  and on the parameters $\sigma^2$ and $\lambda$. As a consequence the resulting process is not a skew Gaussian process.

\begin{figure}[t!]
	\centering
	\caption{Correlation functions for the inline (empty symbols) and circular (solid symbols) process with $\sigma^2=1$, $\delta=0.95$ and exponential correlation function for the processes $X(\mathbf{s})$ and $W(\mathbf{s})$ with  respectively decays parameters 0.5 and 0.5 (circle), 0.5 and 0.2 (triangle), 0.2 and 0.5 (rhombus), 0.2 and 0.2 (square)
	} \label{fig:bicorskewVS}
	\includegraphics[scale=0.5]{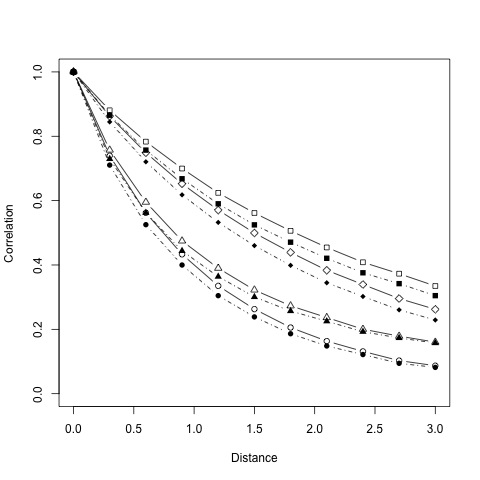}
\end{figure}

The correlation in each of the $X(\mathbf{s})$ and $W(\mathbf{s})$ processes induces association for the $\Theta(\mathbf{s})$ process.
However, because circular variables have no \emph{magnitude} (they only acquire a numerical value given an orientation), there
is no unique way to define the correlation between two circular variables $\Theta(\mathbf{s})$  and
$\Theta(\mathbf{s}^{\prime})$.  A common choice, which exhibits most of the desirable properties of a correlation, is
the one proposed by \cite{Jammalamadaka1988}, that is,
\begin{equation} \label{eq:rhoc}
	\frac{E[\cos(\Theta(\mathbf{s})-\Theta(\mathbf{s}^{\prime})|\boldsymbol{\Psi}) - \cos( \Theta(\mathbf{s})+\Theta(\mathbf{s}^{\prime})+ 2 \tilde{\mu}  |\boldsymbol{\Psi} ) ]  }{    2 \sqrt{   E(\sin^2(\Theta(\mathbf{s}) -\tilde{\mu})|\boldsymbol{\Psi})  E(\sin^2(\Theta(\mathbf{s}^{\prime}) -\tilde{\mu})|\boldsymbol{\Psi})   }   }.
\end{equation}
In our setting (\ref{eq:rhoc}) is not a valid correlation function; it is not a positive definite function. Moreover, we cannot compute  \eqref{eq:rhoc} in closed form but, again, we can resort to Monte Carlo approximation. Figure
\ref{fig:bicorskewVS} provides an illustrative display of the inline and corresponding circular correlations arising from the exponential correlation functions
$\boldsymbol{\rho}_x (h;\boldsymbol{\psi}_x) = \exp(-h \psi_x)$  and $\boldsymbol{\rho}_w (h;\boldsymbol{\psi}_w) = \exp(-h
\psi_w)$.

\subsection{Implementation and Kriging} \label{sec:krig}
%
%
%
%
%
%
Working directly with the wrapped skew Gaussian process is not feasible since  the likelihood for a $n-$dimensional realization of the circular process involves $n$ doubly
infinite sums, i.e. let $\boldsymbol{\Theta}=(\Theta(\mathbf{s}_1),$ $ \Theta(\mathbf{s}_2), \dots , \Theta(\mathbf{s}_n))^{\prime}$ and $\mathbf{K}=(K(\mathbf{s}_1),$  $ K(\mathbf{s}_2), \dots , K(\mathbf{s}_n))^{\prime}$, the density of $\boldsymbol{\Theta}|\boldsymbol{\psi}$ is
\begin{equation}
f(\boldsymbol{\theta}|\boldsymbol{\Psi}) = \sum_{k_1 \in \mathbb{Z}} \sum_{k_2 \in \mathbb{Z}} \dots \sum_{k_n \in \mathbb{Z}} g(\boldsymbol{\theta}+2 \pi \mathbf{k}|\boldsymbol{\Psi}),
\end{equation}
where  $g(\cdot|\boldsymbol{\Psi})$ is the density of $\mathbf{Z}= \boldsymbol{\Theta}+2 \pi \mathbf{K}$, the realization of the skew Gaussian process.     When dealing with wrapped distributions  the winding numbers  are treated as latent random variables \citep[see ][for details and ideas]{Jona2013, coles98}.  Hence,
the joint  distribution of the circular variables and the winding numbers coincides with the joint distribution of the associated
linear variables, i.e., $g(\cdot|\boldsymbol{\Psi})$, and we can work directly with the process $Z(\mathbf{s})$.\\
A critical point is the following.  To simplify the model fitting, recalling \eqref{eq:cond} and \eqref{eq:cond2} and extending them to  $n-$variate random variables, $\mathbf{Z}| \mathbf{X},\boldsymbol{\Psi}$ is normal, hence
the process $Z(\mathbf{s})|X(\mathbf{s}),\boldsymbol{\Psi}$ is Gaussian and $	\Theta(\mathbf{s})|X(\mathbf{s}),\boldsymbol{\Psi}$ is wrapped Gaussian. This implies
that, in the model fitting, if we further introduce the realization of the latent Gaussian process, $X(\mathbf{s})$, along with the set of winding numbers, the $K(\mathbf{s}_{i})$s, then the MCMC implementation  follows directly from the work of
\cite{Jona2013} on the wrapped Gaussian process.\\
In this setting, kriging  is straightforward.
More precisely, let $\mathbf{s}_0$ be the spatial location where we want to predict the circular process and  let $\mathbf{X}=(X(\mathbf{s}_1), X(\mathbf{s}_2), \dots ,$ $ X(\mathbf{s}_n))^{\prime}$. As is customary in the Bayesian framework, to  perform kriging we draw samples from the
predictive distribution of $\Theta(\mathbf{s}_0)|\boldsymbol{\Theta}$:
\begin{align}
f(\Theta(\mathbf{s}_0)|\boldsymbol{\Theta}) = &
\sum_{\mathbf{K}\in \mathbb{Z}^n} \int_{\boldsymbol{\Psi}} f(\Theta(\mathbf{s}_0)|X(\mathbf{s}_0),\mathbf{X}, \mathbf{K},\boldsymbol{\Psi},\boldsymbol{\Theta}) \times  \\ &f(X(\mathbf{s}_0)|\mathbf{X},\boldsymbol{\Psi})f(\mathbf{X}, \mathbf{K},\boldsymbol{\Psi}|\boldsymbol{\Theta} ) d \boldsymbol{\Psi}.\label{eq:krig}
\end{align}
Let $\boldsymbol{\Psi}^b$, $\mathbf{X}^b$ and $\mathbf{K}^b$ be the $b^{th}$ sample from the posterior distribution $f(\mathbf{X}, \mathbf{K},\boldsymbol{\Psi}|\boldsymbol{\Theta} )$.
We can  sample from \eqref{eq:krig} with composition sampling.  That is, if for each posterior sample we simulate $X^{b}(\mathbf{s}_{0})$ from the distribution  $X(\mathbf{s}_0)|\mathbf{X}^b,\boldsymbol{\Psi}^b$ and $\Theta^{b}(\mathbf{s}_{0})$ from the distribution  $\Theta(\mathbf{s}_0)|X^{b}(\mathbf{s}_{0}),$ $\mathbf{X}^b, \mathbf{K}^b,$ $\boldsymbol{\Psi}^b,\boldsymbol{\Theta}$, then each $\Theta^{b}(\mathbf{s}_{0})$ can be considered as a sample from \eqref{eq:krig}. \\We can easily simulate $X^{b}(\mathbf{s}_{0})$ since
$X(\mathbf{s}_0),\mathbf{X}^b|\boldsymbol{\Psi}^b$ is Gaussian  and then   $X(\mathbf{s}_0)|\mathbf{X}^b,\boldsymbol{\Psi}^b$      is univariate normal with mean and covariance that can be derived using standard results.  If we simulate $Z^{b}(\mathbf{s}_{0})$ from $Z(\mathbf{s}_0)|\mathbf{Z}^b,$ $X^{b}(\mathbf{s}_{0}),\mathbf{X}^b ,\boldsymbol{\Psi}^b$, where $\mathbf{Z}^b = \boldsymbol{\Theta}+2 \pi \mathbf{K}^b$,  we can immediately obtain $\Theta^{b}(\mathbf{s}_{0})$ as  $\Theta^{b}(\mathbf{s}_{0})= Z^{b}(\mathbf{s}_{0}) \mbox{ mod }2 \pi$, that is a sample from  $\Theta(\mathbf{s}_0)|X^{b}(\mathbf{s}_{0}),$  $\mathbf{X}^b,  \mathbf{K}^b,\boldsymbol{\Psi}^b,\boldsymbol{\Theta}$. Remark that to obtain a sample of  $Z^{b}(\mathbf{s}_{0})$  is really easy since
\begin{align}
&\left(
\begin{array}{c}
Z(\mathbf{s}_0)\\
\mathbf{Z}
\end{array}
\right) | \mathbf{X},X(\mathbf{s}_0),\boldsymbol{\Psi}
\sim \\
& N \left(
\begin{array}{c}
\mu^*+\frac{\sigma \lambda}{\sqrt{1+\lambda^2}}|X(\mathbf{s}_0)|\\
\mu^*\mathbf{1}_n+\frac{\sigma \lambda}{\sqrt{1+\lambda^2}}|\mathbf{X}|
\end{array} , \frac{\sigma^2}{1+\lambda^2}
\left(
\begin{array}{cc}
1 & \boldsymbol{\rho}_{0,w}^{\prime}\\
\boldsymbol{\rho}_{0,w} & \mathbf{\Upsilon}
 \end{array}
\right)
\right)
\end{align}
where  $\mathbf{1}_{n}$ is a vector of 1s of dimension $n$,   $(\mathbf{\Upsilon})_{ij}=\rho_{w}(|| \mathbf{s}_i-\mathbf{s}_j ||;\boldsymbol{\psi}_w)$ and $(\boldsymbol{\rho}_{0,w})_{i}= \rho_{w}(|| \mathbf{s}_i-\mathbf{s}_0 ||;\boldsymbol{\psi}_w)$. Then the  distribution of $Z(\mathbf{s}_0)|\mathbf{Z}^b,X^{b}(\mathbf{s}_{0}),\mathbf{X}^b ,\boldsymbol{\Psi}^b$  is normal. \\


\section{A dynamic extension of the wrapped skew Gaussian process} \label{sec:spattemp}


We extend our model to the dynamic setting following ideas in \cite{Banerjee2014}. We start by specifying an inline process $Z_t(\mathbf{s})$, $t \in [1,\dots, T]$,
as
\begin{align}
	Z_1(\mathbf{s}) & = \mu+  \frac{\sigma \lambda}{\sqrt{1+\lambda^2}  }   |X_1(\mathbf{s})|\\
	&+  \frac{\sigma}{\sqrt{1+\lambda^2}  } W_1(\mathbf{s})-  \frac{\sigma \lambda \sqrt{ 2  }}{ \sqrt{\pi(1+\lambda^2)}  },\label{eq:dynamiclinear} \\
	Z_t(\mathbf{s}) & = \mu +\gamma (Z_{t-1}(\mathbf{s})-\mu)+ \frac{\sigma \lambda}{\sqrt{1+\lambda^2}  }   |X_t(\mathbf{s})|\\
	&+  \frac{\sigma}{\sqrt{1+\lambda^2}  } W_t(\mathbf{s})-  \frac{\sigma \lambda \sqrt{ 2  }}{ \sqrt{\pi(1+\lambda^2)}  },\,  t \neq 1, \label{eq:dynamiclinear2}
\end{align}
where $\gamma \in [-1,1]$,  $\forall t$ we have $X_t(\mathbf{s})|\boldsymbol{\Psi} \sim GP (0, \boldsymbol{\rho}_x (h;\boldsymbol{\psi}_x) ) $,
$W_t(\mathbf{s}) |\boldsymbol{\Psi}\sim GP (0, \boldsymbol{\rho}_w (h;\boldsymbol{\psi}_w) ) $ and
$\mbox{Cov}(X_t(\mathbf{s}),X_{t^{\prime}}(\mathbf{s}^{\prime})|\boldsymbol{\Psi})=\mbox{Cov}(W_t(\mathbf{s}),W_{t^{\prime}}(\mathbf{s}^{\prime})|\boldsymbol{\Psi})
= 0$ if $t \neq t^{\prime}$.  Expressions \eqref{eq:dynamiclinear} and \eqref{eq:dynamiclinear2} provide a mean-centered, first order auto-regressive model
with i.i.d. process increments.  Moreover, the process increments are skew GP's with parameters $\sigma, \lambda, \rho_{x},
\rho_{w}$. Equivalently, we see that $Z_1(\mathbf{s})|\boldsymbol{\Psi} \sim SGP(\mu, \sigma^2, \lambda)$ and $Z_t(\mathbf{s}) |
Z_{t-1}(\mathbf{s}) ,\boldsymbol{\Psi}\sim SGP(\mu+\gamma(Z_{t-1}(\mathbf{s})-\mu), \sigma^2, \lambda)$.
%
%

Under the dynamic spatial setting, we are generally interested in predicting the process (i) at an  observed spatial location at
time $T+h$, $h \in \mathbb{Z}^+$ (usually $h=1$) or (ii) at an unobserved spatial location  $\mathbf{s}_0$ inside the observed time window.
Suppose we let $\mu^b$, $(\sigma^2)^b$, $\lambda^b$ and $\gamma^b$ be the samples of the parameters of the $b^{th}$ iteration of
the MCMC algorithm, $({\mu}^*)^{b} = \mu^b- {\sigma^b \lambda^b \sqrt{ 2  }}/{ \sqrt{\pi(1+(\lambda^b)^2)}  }$, $X_{t}^b(\mathbf{s})$ and $K_{t}^b(\mathbf{s})$ the $b^{th}$ realization of the  processes
$X_{t}(\mathbf{s})$ and $K_{t}(\mathbf{s})$ at site $\mathbf{s}$  and time $t$ and $Z_{t}^b(\mathbf{s}) = X_t(\mathbf{s})+2 \pi
K_t^b(\mathbf{s})$.
$B$ samples from the predictive distribution $\Theta_{T+h}(\mathbf{s})| \boldsymbol{\Theta} $, where
$\boldsymbol{\Theta}$ is the observed circular data,  can be obtained if, for each MCMC sample, we draw a value
$Z_{T+h}^b(\mathbf{s})$ from a normal distribution with mean
$$
({\mu}^*)^{b} +(\gamma^b)^h(Z_{T}^b(\mathbf{s})-\mu^b)+\frac{\sigma^b \lambda^b}{\sqrt{1-(\lambda^b)^2}}   |X_{T+h}^b(\mathbf{s})|
$$
and variance
$$
\frac{(\sigma^2)^b}{1-(\lambda^b)^2}.
$$
The set $\{\Theta_{T+1}^b(\mathbf{s})\}_{b=1}^B$ is from the desired predictive
distribution.

To obtain the $b^{th} $ posterior sample of the predictive distribution of  $\Theta_t(\mathbf{s}_0)|\boldsymbol{\Theta}$ we adopt the
usual composition sampling by first sampling $X_t^b(\mathbf{s}_0)$ from the distribution of $X_t(\mathbf{s}_0)|\mathbf{X},\boldsymbol{\Psi}^b$ and
then sampling $Z_t^b(\mathbf{s}_0)$ from $Z_t(\mathbf{s}_0)|\mathbf{Z},\mathbf{X},X_t^b(\mathbf{s}_0),\boldsymbol{\Psi}^b$. Finally,
$\Theta_t^b(\mathbf{s}_0) =Z_t^b(\mathbf{s}_0) \mbox{ mod } 2 \pi $ is a draw from the  predictive distribution
$\Theta_t(\mathbf{s}_0)|\boldsymbol{\Theta}$.

The distribution of  $Z_t(\mathbf{s}_0),\mathbf{Z}|\mathbf{X},X_t^b(\mathbf{s}_0),\boldsymbol{\Psi}^b$ is again multivariate normal and for
spatial locations $\mathbf{s}_{i}, i=1,2,...,n$, let $\mathbf{Z}_t= (Z_t(\mathbf{s}_1),Z_t(\mathbf{s}_2), \dots ,
Z_t(\mathbf{s}_n))^{\prime}$, $\mathbf{Z} =(\mathbf{Z}_1,\mathbf{Z}_2,\dots , \mathbf{Z}_T)^{\prime}$ and $\mathbf{X}$  be the
associated realization of the process $X(\mathbf{s})$.  Let $\boldsymbol{\Gamma}$ be a $T \times T$ correlation matrix with
$i,j$th element  equal to $\gamma^{|i-j|}$, $\boldsymbol{\Gamma}_l$ be the lower triangular part of $\boldsymbol{\Gamma}$  and
$\mathbf{C}$  be the correlation matrix of $\mathbf{W}_t =(W_t(\mathbf{s}_1),W_t(\mathbf{s}_2), \dots ,
W_t(\mathbf{s}_n))^{\prime}$. Let $\mathbf{D}$ be a vector of length $n$ with $i^{th}$ element equal to
$\mbox{Cor}(W_t(\mathbf{s}_0),W(\mathbf{s}_i))$, $\mathbf{F}_t$ be a vector of length $T$ with $i^{th}$ element equal to
$\gamma^{|t-i|}$, $\mathbf{I}_n$ be the identity matrix of dimension $n$ and
let $\otimes$ indicates the Kronecker product. Altogether, we have that
\begin{align}
	&\left(\begin{array}{c}
		Z_t(\mathbf{s}_0)\\
		\mathbf{Z}
	\end{array}\right)| \mathbf{X},X_t(\mathbf{s}_0),\boldsymbol{\Psi} \sim \\
	&N
	\left(
	\begin{array}{c}
		{\mu}^* +\frac{\sigma \lambda}{\sqrt{1-\lambda^2}}   |X_t(\mathbf{s}_0)|   \\
		\boldsymbol{\delta}
	\end{array},\frac{\sigma^2}{1-\lambda^2}
	\left(
	\begin{array}{cc}
		1 & (\mathbf{F}_t \otimes \mathbf{D})^{\prime}\\
		\mathbf{F}_t \otimes \mathbf{D} & \boldsymbol{\Gamma} \otimes \mathbf{C}
	\end{array}
	\right)
	\right)
\end{align}
where
\begin{align}
\boldsymbol{\delta}= \mu \mathbf{1}_{nT}&+{\sigma \lambda}/{\sqrt{1-\lambda^2}}   \left(  \boldsymbol{\Gamma}_l \otimes \mathbf{I}_n
\right)|\mathbf{X}|\\
&-   {\sigma \lambda \sqrt{ 2  }}/{ \sqrt{\pi(1+\lambda^2)}  }   \left(  \boldsymbol{\Gamma}_l \otimes \mathbf{I}_n \right)\mathbf{1}_{nT}.
\end{align}
 Here, again
$Z_t(\mathbf{s}_0)|\mathbf{Z},\mathbf{X},X_t^b(\mathbf{s}_0),\boldsymbol{\Psi}^b$ is univariate normal and sampling from it is easy.

\section{A brief simulation study} \label{sec:sim}

\begin{figure}[t!]
	\centering
	\caption{Spatial correlation functions for the simulated datasets: circles are associated to Data1 ($\lambda=0$), triangles to  Data2 ($\lambda=1.5$), diamonds to Data3 ($\lambda=3$) and squares to  Data4 ($\lambda=10$).
	} \label{fig:corSkew}
	\includegraphics[scale=0.5]{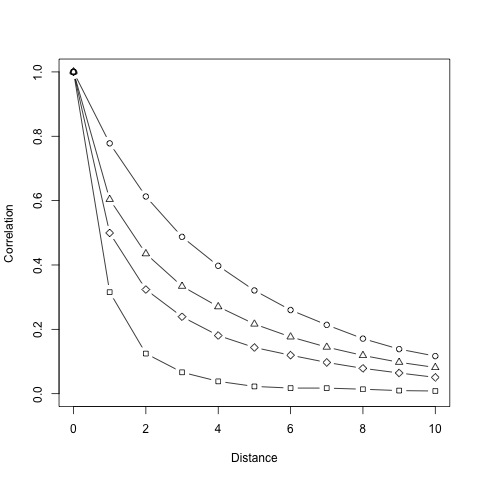}
\end{figure}

 We fit and estimate the model proposed in   Section \ref{sec:spattemp} to 8 datasets simulated with $\mu=\pi$,  $\sigma^2=1$ and  4 levels of the skew parameter  $ \lambda = \{0.0,1.5, 3  , 10 \}$. For  the AR(1) parameter we chose $\gamma= 0.5$; we experimented with several values of $\gamma\in(0,1)$ obtaining similar results, so we report  estimates using the central value of the interval. We  work with  2
	sample sizes, 110 spatial locations and 60 time points, ($N=n\times T=110\times 60$), 220 spatial locations and 60  time points, $N=220\times 60$), to assess whether there are differences in the parameter estimates when the sample size increases.
The coordinates are uniformly generated over $[0,10]^2$ and for both processes, $X_t(\mathbf{s})$  and
$W_t(\mathbf{s})$, we adopt exponential correlation functions. We choose $\psi_x = 0.5$ and $\psi_w= 0.2$ and notice that, as $\lambda$ varies, we obtain  different
spatial correlations as shown in  Figure \ref{fig:corSkew}.
 
The model is estimated with  90\% of the spatial locations, i.e. 100 for the first sample size and 200 for the second, using the first 50 time points.  Therefore, the training set is made of $100\times50$ and $200\times50$ points.  We select observations using simple random sampling on the spatial locations (probability of inclusion in the training set $1/n$). The remaining spatial locations and 10 final time points are used to define two types of validation sets: (i) prediction at observed times, i.e. we use observations between time 1 and time 50 not used to estimate the models. To simplify we call this set the \emph{spatial validation set}; (ii) prediction at unobserved times, i.e. we use observations from time 51 to time 60 at   spatial locations used to estimate the models. We call this set the \emph{ temporal validation set.} We repeat the sampling procedure 40 times.

%
%
%
%
%
%
%
%
%

As prior distributions we use $\mu \sim U(0,2\pi)$, $\gamma \sim U(-1,1)$, $\psi_x\sim U(0.1,1)$ and $\psi_w\sim U(0.1,1)$. To
choose the prior on $\sigma^2$ and $\lambda$ we note that, as for the wrapped Normal case \citep{Jona2013}, if  the variance of
the associated inline distribution increases we are unable to tell the difference between the wrapped skew normal and a circular uniform. As we noted in Section \ref{sec:unv},  the variance of the skew normal is  $${\sigma^2 \lambda^2}/{( {1+\lambda^2})  }  \left(1-{2}/{\pi}\right)+ {\sigma^2 }/{ ({1+\lambda^2})  },   $$ i.e., it is a
function of both $\sigma^2$ and $\lambda$.  In this regard, when  $\sigma^2=10$, with sample size of 200, independently of
$\lambda$, the Rayleigh test of (circular) uniformity  fails to  discriminate between the wrapped skew normal and the circular uniform. So, we chose $\sigma^2 \sim U(0,10)$ and a weak informative prior for $\lambda$, $\lambda \sim N(0,100)$.

\begin{table*}[t!]
	\centering
	$n=110$ \\
	\begin{tabular}{c|cccc}
		\hline& Data1 ($\lambda=0$) & Data2 ($\lambda=1.5$)& Data3 ($\lambda=3$)& Data4 ($\lambda=10$) \\ \hline \hline
		$\hat{\mu}$  & 3.03 & 3.365&3.217  & 3.109\\
		C.I.&(2.762 3.321) &(3.205 3.533)  & (3.106 3.334) &(3.044 3.177)\\
		$\hat{\sigma}^2$   & 1.715&1.213 &1.061  & 0.962 \\
		C.I.&(1.390 2.186) &(1.080 1.388)  & (0.976 1.177) &(0.888 1.046) \\
		$\hat{\lambda}$  & 0.931& 1.690& 3.278 &9.864 \\
		C.I.&(0.689 1.275) &(1.498 1.924)  & (2.881 3.716) &(8.572 11.282)\\
		$\hat{\gamma}$  &0.388 & 0.446&0.499  & 0.488 \\
		C.I.&(0.35  0.42) &(0.421 0.470)  & (0.479 0.518) &(0.475 0.502) \\
		$\hat{\psi}_x$  &0.234 &0.399 & 0.472 & 0.528 \\
		C.I. &(0.139 0.483) &(0.332 0.473)  & (0.413 0.528) &(0.475 0.589)\\
		$\hat{\psi}_w$  &0.144 &0.254 &  0.191& 0.210 \\
		C.I.&(0.109 0.186) &(0.195 0.318)  & (0.141 0.251) &(0.137 0.307)\\
		\hline \hline		
	\end{tabular}
\vspace{0.2cm} \\
	\centering
	$n=220$\\
	\begin{tabular}{c|cccc}
		\hline& Data1 ($\lambda=0$) & Data2 ($\lambda=1.5$)& Data3 ($\lambda=3$)& Data4 ($\lambda=10$) \\ \hline \hline
		$\hat{\mu}$  &2.981  & 3.353& 3.209 &3.094 \\
		C.I.&(2.713 3.261) &(3.209 3.504)  & (3.067 3.346) &(3.031 3.161)\\
		$\hat{\sigma}^2$   & 1.448& 1.087& 1.097 & 0.956 \\
		C.I.&(1.266 1.701) &(0.994 1.196)  & (1.005 1.211) &(0.887 1.034) \\
		$\hat{\lambda}$  &-0.716 & 1.383& 2.501 &9.619 \\
		C.I.&(-0.869 -0.589) &(1.242 1.532)  & (2.227 2.777) &(8.449 10.771)\\
		$\hat{\gamma}$  &0.370 &0.436 & 0.488 &0.499 \\
		C.I.&(0.349 0.390) &(0.418 0.452)  & (0.474 0.503) &(0.490 0.507) \\
		$\hat{\psi}_x$  & 0.430&0.558 &0.500  & 0.511 \\
		C.I. &(0.323 0.625) &(0.485 0.639)  & (0.444 0.558) &(0.467 0.555)\\
		$\hat{\psi}_w$  &0.152 &0.286 &  0.192&0.152 \\
		C.I.&(0.119 0.186) &(0.235 0.340)  & (0.143 0.245) &(0.112 0.212)\\
		\hline \hline		
	\end{tabular}
	\caption{Parameter estimates (mean) and credible intervals (C.I.) for the wrapped skew Gaussian model in the 4 simulated datasets} \label{tab:estsim2}
\end{table*}

\begin{table*}[t!]
	\centering
	$n=110$\\
	\begin{tabular}{c|cccc}
		\hline& Data1 ($\lambda=0$) & Data2 ($\lambda=1.5$)& Data3 ($\lambda=3$)& Data4 ($\lambda=10$) \\ \hline \hline
		$\hat{\mu}$  & 2.986 & 3.313 & 3.208 &3.138 \\
		C.I.&(2.752 3.222) &(3.211 3.409)  & (3.123 3.290) &(3.082 3.199) \\
		$\hat{\sigma}^2$  &1.141  & 0.596 & 0.465 & 0.369 \\
		C.I.&(0.993 1.340) &(0.556 0.645)  & (0.438 0.497) &(0.35  0.39)\\
		$\hat{\gamma}$ & 0.415 & 0.417 & 0.489 &0.488 \\
		C.I.&(0.388 0.441) &(0.392 0.441)  & (0.465 0.514) &(0.463 0.514) \\
		$\hat{\psi}_w$  &0.225  & 0.67  & 0.796 & 1.182 \\
		C.I. &(0.189 0.261) &(0.611 0.726)  & (0.731 0.862) &(1.099 1.265)\\ \hline \hline
	\end{tabular}
\vspace{0.2cm} \\
	\centering
	$n=220$\\
	\begin{tabular}{c|cccc}
		\hline& Data1 ($\lambda=0$) & Data2 ($\lambda=1.5$)& Data3 ($\lambda=3$)& Data4 ($\lambda=10$) \\ \hline \hline
		$\hat{\mu}$  & 3.023 &  3.308& 3.181 & 3.143\\
		C.I.&(2.833 3.210) &(3.216 3.403)  & (3.103 3.254) &(3.090 3.205) \\
		$\hat{\sigma}^2$  &1.061  & 0.602  & 0.473 & 0.370  \\
		C.I.&(0.937 1.209) &(0.564 0.647)  & (0.449 0.501) &(0.354 0.390)\\
		$\hat{\gamma}$ &0.365  &0.426  &0.468  & 0.503 \\
		C.I.&(0.346 0.384) &(0.407 0.444)  & (0.452 0.487) &(0.486 0.519) \\
		$\hat{\psi}_w$  &0.273  & 0.678 & 0.867 & 1.152 \\
		C.I. &(0.237 0.309) &(0.626 0.730)  & (0.809 0.923) &(1.081 1.218)\\ \hline \hline	
	\end{tabular}
	\caption{Parameter estimates (mean) and credible intervals (C.I.) for the wrapped  Gaussian model in the 4 simulated datasets} \label{tab:estsim4}
\end{table*}

%

%
%

For each dataset we also fit a wrapped normal model (setting $\lambda=0$) and we compare the models with regard to posterior
point estimates and predictive  ability. 
The predictive ability of the models is evaluated  by computing the continuous
rank probability score (CRPS)  for circular variables \citep{grimit2006}. The CRPS is a proper scoring rules  defined, for
circular variables,  as
\begin{equation} \label{eq:crps}
	CRPS(F,  \xi)= E(d(\Xi,\xi))-\frac{1}{2}E(d(\Xi,\Xi^*)),
\end{equation}
where $F$ is the predictive distribution, $\xi$ is the holdout value, $\Xi$  and $\Xi^*$ are independent copies of a
circular variable with distribution $F$, and $d(\Xi, \Xi^*) = 1- {\cos}(\Xi - \Xi^*)$, the circular distance
\citep[p. 15]{Jammalamadaka2001}.
Exact calculation of \eqref{eq:crps} is not possible since we can not obtain the predictive distribution under the skew or the non
skew Gaussian process in closed form.  However, for the validation point $\theta_t(\mathbf{s}_0)$ we can compute a Monte Carlo approximation
as
\begin{equation}
	\frac{1}{B}\sum_{b=1}^Bd(\theta_t^b(\mathbf{s}_0),\theta_t(\mathbf{s}_0))-\frac{1}{2B^2}\sum_{l=1}^B\sum_{b=1}^B d(\theta_t^l(\mathbf{s}_0),\theta_t^b(\mathbf{s}_0)) \label{eq:crps1}
\end{equation}
where $\theta_t^b(\mathbf{s}_0)$ denotes the simulated value of $\theta_t(\mathbf{s}_0)$ using the $b^{th}$ posterior
parameters and $B$ is the total number of posterior samples.

\begin{figure}[t!]
	\centering
	\caption{Illustrative predictive distributions for a holdout site in the first  (a) and in the fourth simulated dataset (b). The solid line is the predictive distribution under the wrapped Gaussian model while the dashed one is under the wrapped skew Gaussian model. The vertical line represents the true holdout simulated  value} \label{fig:simpred1}
	\subfloat[Data1]{\includegraphics[scale=0.35]{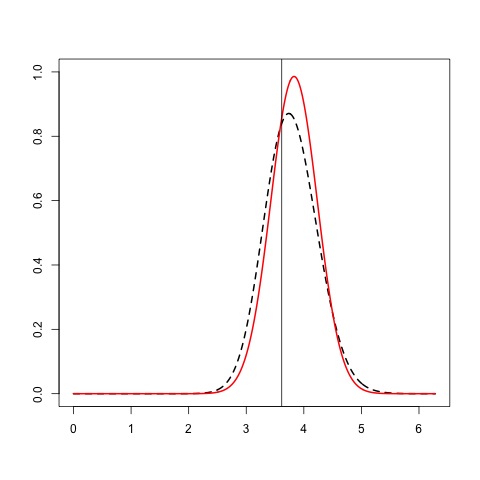}}\\
	\subfloat[Data4]{\includegraphics[scale=0.35]{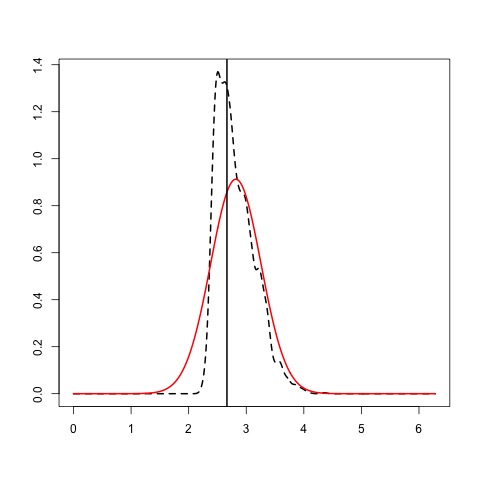}}
\end{figure}

As an example, in Tables  \ref{tab:estsim2} and
\ref{tab:estsim4} we present the posterior  mean estimates and credible intervals for all the parameters in all simulated
datasets using one training set, i.e. the same locations and times for each dataset.
For the fourth dataset and for both sample sizes, the
skew model  well estimates the parameters (the true value is inside the credible interval (C.I.)). In the first dataset $\lambda$ is far from 0.
The wrapped skew normal process shows a substantial gain relative to the wrapped Gaussian process in terms of predictive
ability for locations inside the observed time windows, even if the true model used to simulate the data is the wrapped Gaussian
(Data1), see Table \ref{tab:crpssim2}. As for forecasting (temporal validation set), we see that there is no difference between
the models in terms of CPRS. Illustrative comparison of the predictive distributions under the two models can be seen in Figure \ref{fig:simpred1}.   As we expect,in the fourth dataset the predictive distribution is  highly skewed while, in the first, it is essentially  symmetric.

\begin{table}[t!]
	\centering
	Spatial\\
	\begin{tabular}{c|c|cccc}
		\hline&& Data1 & Data2 & Data3 & Data4 \\ \hline \hline
		n=110 & WS & 0.173 & 0.146 & 0.118  &  0.116 \\
		& W & 0.221 & 0.179 & 0.176 & 0.160 \\ \hline
		n=220 & WS & 0.170 & 0.149 & 0.116 & 0.091  \\
		& W & 0.205 & 0.179 & 0.168  &  0.148 	\\ \hline \hline	
	\end{tabular}
	\vspace{0.2cm}\\
	Temporal		\\
	\begin{tabular}{c|c|cccc}
		\hline && Data1 & Data2 & Data3 & Data4 \\ \hline \hline
		n=110 & WS& 0.348 & 0.266 & 0.188  & 0.181  \\
		& W &   0.349  & 0.265 &0.191 & 0.184 \\ \hline
		n=220 & WS&  0.350 & 0.275 &0.193  & 0.181  \\
		& W & 0.349 & 0.272 & 0.194 & 0.183	\\ \hline \hline	
	\end{tabular}
	\caption{Simulated datasets: mean CRPSs over 40 validation sets. Models based on the wrapped skew normal (WS) and the wrapped normal (W)}\label{tab:crpssim2}
\end{table}

\section{The wave direction data example}

\begin{table*}[t!]
	\centering
	\begin{tabular}{c|cc|cc}\hline
		& calm & calm & storm & storm \\
		& WS & W & WS & W \\ \hline \hline
		$\hat{\mu}$  &3.372    & 3.19 & 3.398 &3.39 \\
		C.I.& (2.610  4.150)  & (2.905 3.500) &(2.498 4.274) & (2.939 3.851)  \\
		$\hat{\sigma}^2$   & 5.246 & 1.827 & 5.015   & 1.283\\
		C.I.  &(4.214 6.883) & (1.526 2.276) & (4.029 6.581)  & (1.130 1.477)  \\
		$\hat{\lambda}$ & 1.432 & $\cdot$ &1.159 & $\cdot$ \\
		C.I. &(1.068 1.762) &  ($\cdot$ $\cdot$) &     (0.868 1.496)  & ($\cdot$ $\cdot$)\\
		$\hat{\gamma}$&  0.438  & 0.567 & 0.377 &0.479\\
		C.I.& (0.406 0.471)  & (0.540 0.594) &(0.350 0.406) & (0.453 0.504)  \\
		$\hat{\psi}_x$ &0.006 &  $\cdot$ &  0.006& $\cdot$\\
		C.I.&     (0.005 0.008)  & ($\cdot$ $\cdot$) &(0.005 0.008) & ($\cdot$ $\cdot$) \\
		$\hat{\psi}_w$ &  0.002  & 0.013 & 0.001 &0.007\\
		C.I. & (0.001 0.003)  & (0.011 0.015) &(0.001 0.001) & (0.005 0.008) \\ \hline \hline		
	\end{tabular}
	\caption{Parameter estimates (mean) and credible intervals (C.I.) for the wave direction data} \label{tab:estreal}
\end{table*}

The real data we use come from a deterministic wave model implemented by Istituto Superiore per la Protezione e la Ricerca Ambientale (ISPRA) that gives hourly  prediction over a
grid of about 12.5$\times$12.5 Km on the Adriatic sea \citep{speranza2004}. Over the Adriatic Sea area, there are 1494 points, with minimum and maximum distance of about 7km and 852km respectively.
The computer model starts from a wind forecast model predicting the surface wind over the entire Mediterranean and then the prediction of the  wave direction is obtained
solving energy transport equations using the wind forecast as input.\\

We developed two datasets. The first spans the period  April 2010 between the 2nd at 00:00 and the 4th at 22:00, a calm period.  The second spans  the period  April 2010 between the 5th at 00:00 and the 7th at
22:00, a storm period. 
We randomly  select  220 spatial locations; the same spatial locations are used for the calm and storm period dataset.\\

Similarly to what we did in the simulated examples, we use 90\% of the spatial locations, taking the first  48 time points to estimate models while  the remaining locations and times are included in the building of the two types of validation sets. \\

Again, for
each training set, we fitted a  skew Gaussian model  and a wrapped Gaussian model. We repeat the splitting procedure into  training and validation sets 40 times and each time we compute the CRPS to compare the performance of the models. \\
As prior distributions we used the same choices as in Section \ref{sec:sim}
with the exception of the spatial decays;
for $\psi_w$ we adopt a $U(10^{-3},10^{-1})$  which corresponds to a maximum and minimum practical  range of 3000km and 30km while for $\psi_x$ we adopt a $U(5^{-4},5^{-2})$ which roughly corresponds to the same practical spatial range for the  process $|X(\mathbf{s})|$.

\begin{table}[t!]
	\centering
	\begin{tabular}{c|cc|cc}\hline
		& calm & calm & storm & storm 	\\
		& WS & W & WS & W \\ \hline \hline
		Spatial   &     0.426 & 0.494  & 0.528  & 0.567 \\ \hline \hline
		Temporal  & 0.520 & 0.628  &0.446  & 0.476	     	\\ \hline	
	\end{tabular}
	\caption{{Wave data:  mean CRPSs over 40 validation sets. Models based on the wrapped skew normal (WS) and the wrapped normal (W)}} \label{tab:crpsreal}
\end{table}

\begin{figure}[t!]
	\centering
	\caption{Examples of predictive distributions for one of the holdout site in calm  (a) and storm (b) sea state. The solid line is the predictive distribution under the wrapped Gaussian model while the dashed one is under the skew Gaussian model. The vertical line represents the true holdout observed value} \label{fig:simpredReal}
	\subfloat[Calm]{\includegraphics[scale=0.35]{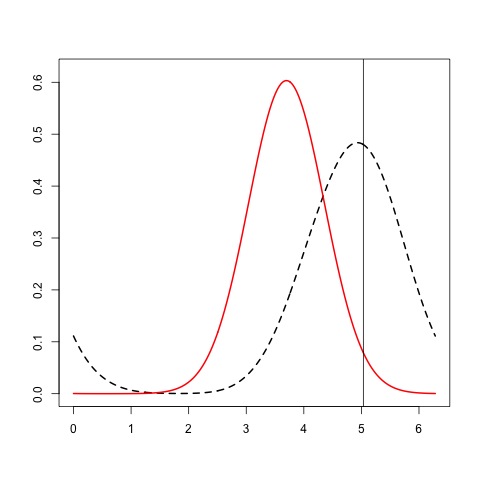}}\\
	\subfloat[Storm]{\includegraphics[scale=0.35]{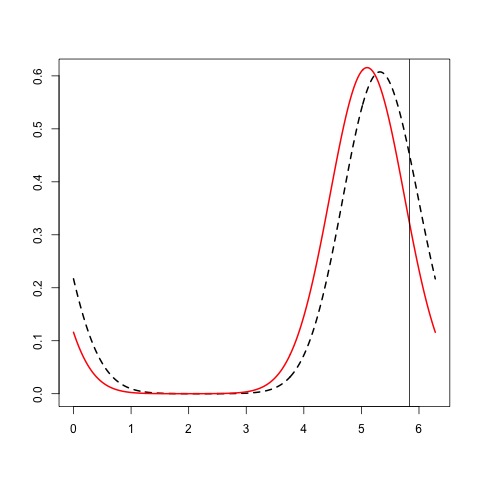}}
\end{figure}

In Table \ref{tab:estreal} we provide the parameters estimates for the first selected training sets. The  estimated spatial dependence ($\psi_w$) of the $W(\mathbf{s})$ process is stronger during the storm for both models while ($\psi_x$) seems to remain the same in both sea states for $X(\mathbf{s})$. 
Again, employing the CRPS, for both validation sets under both sea states, the wrapped skew Gaussian process shows a consequential gain in
predictive ability compared with the standard wrapped Gaussian.\\
Finally, Figure \ref{fig:simpredReal}  shows   examples of predictive distributions for a holdout sample during a calm and a storm state. We showed in Figure \ref{fig:skew1} that  with $|\lambda|<3$ there is little difference between the (symmetric) wrapped normal and the (asymmetric) wrapped skew normal. Since, in these two examples  $|\hat{\lambda}|<1.5$,  the  predictive distributions under the skew normal models are  roughly symmetric.

\section{Summary and future work}

We have presented a novel process model for dynamic spatial directional data.  That is, we have a conceptual time series of
directions at each spatial location in the region and we observe these series for a finite collection of locations. The model,
referred to as a wrapped skew Gaussian process, enables more flexible marginal distributions for the locations than the
symmetric ones that are available under the previously published wrapped Gaussian process.  Using both simulation and a wave
direction dataset, we are able to show improved out-of-sample prediction with the former.

Future work offers several opportunities.  One is to note that wave heights are available in addition to wave directions. Wave
heights inform about the sea state and therefore whether we are in a calm, storm, or transition state.  In particular,
predictive uncertainty varies with wave height and/or sea state, e.g., prediction is more precise during storm. So, we can
attempt to extend the proposed model to introduce covariates into the mean model and also into the variance model for the
wrapped skew Gaussian process. Another possibility is to model temporal data, where the time of the observed event is treated as
random. Then, upon wrapping, we would have circular times.  In addition, the locations of the events are random.  The data would be treated as a point pattern over space
and (circular) time.


\bibliographystyle{spbasic}      
\bibliography{all}   
%

\end{document}